\documentclass[prd,10pt,twocolumn,nofootinbib,preprint,superscriptaddress]{revtex4}
\pdfoutput=1
\usepackage{feynmp}
\usepackage{amsmath, amssymb, amsthm, graphicx, fancyhdr,epsfig, slashed, mathrsfs}
\usepackage{pifont}
\usepackage{subfigure}
\usepackage{tikzsymbols}
\usepackage{natbib}
\usepackage{float}
\usepackage{lipsum}
\usepackage{mathtools} 
\usepackage{setspace}
\usepackage[normalem]{ulem}
\usepackage{tabularx}
\usepackage{tikz,xcolor,hyperref}
\usepackage{cancel}
\hypersetup{
     colorlinks=true,
     citecolor    = blue,
     linkcolor= blue
}

\definecolor{lime}{HTML}{A6CE39}
\DeclareRobustCommand{\orcidicon}{
	\begin{tikzpicture}
	\draw[lime, fill=lime] (0,0) 
	circle [radius=0.2] 
	node[white] {{\fontfamily{qag}\selectfont \tiny ID}};
	\draw[white, fill=white] (-0.0625,0.095) 
	circle [radius=0.007];
	\end{tikzpicture}
	\hspace{-2mm}
}

\foreach \x in {A, ..., Z}{\expandafter\xdef\csname orcid\x\endcsname{\noexpand\href{https://orcid.org/\csname orcidauthor\x\endcsname}
			{\noexpand\orcidicon}}
}
\def\d{\mathrm{d}}
\newcommand{\nuc}{\mathcal{N}}
\newcommand{\qt}{\left|\mathbf{q}\right|}

\begin{document}

\title{Atmospheric neutrino up-scattering explanation of LZ 2026 excess }

\author{Sk Jeesun\orcidA}
\email{jeesun@sjtu.edu.cn}
\affiliation{State Key Laboratory of Dark Matter Physics, Tsung-Dao Lee Institute $\&$ School of Physics and Astronomy, Shanghai Jiao Tong University, Shanghai 200240, China}
\affiliation{Key Laboratory for Particle Astrophysics and Cosmology (MOE) $\&$ Shanghai Key Laboratory for Particle Physics and Cosmology, Shanghai Jiao Tong University, Shanghai 200240, China}
\author{Anirban Majumdar\orcidB}
\email{anirban19@iiserb.ac.in}
\affiliation{Department of Physics, Indian Institute of Science Education and Research - Bhopal,
Bhopal Bypass Road, Bhauri, Bhopal 462066, India}

\begin{abstract}
The recent observation of an isolated nuclear recoil at $248\pm 23\pm 23$ keV energy by the LUX-ZEPLIN (LZ) experiment has motivated the community to look for a new physics explanation, as the Standard Model background estimation fails to accommodate it.
Most of the existing literature hitherto considers a galactic halo dark matter with a heavier partner.
In this work, we traverse the alternate route of atmospheric neutrino ($\nu$) up-scattering, thus producing a massive beyond standard model (BSM) particle $\chi$ with mass $m_\chi$.
The kinematic requirement of such a scattering demands a higher energy of the incoming neutrinos than elastic scattering, explaining the absence of excess events from solar $\nu$ scattering.
However, for high energy neutrinos in the tail of the atmospheric neutrino flux, the kinematic threshold no longer prevents events at low recoil energies.
To avoid an excess of such events, we consider a momentum dependent operator, which suppresses the event rate at low recoil energies and allows a single event in the high recoil energy window.
Such up-scattering on nuclei ($\mathcal{N}$), $\nu \mathcal{N}\to \chi \mathcal{N}$, can be naturally realized in sterile neutrino models, though we keep our analysis generic without specifying $\chi$.
We identify the region of parameter space that can produce such an isolated event, assuming a pseudoscalar mediator with mass $m_\phi$ that couples to the $\nu$-$\chi$ pair with strength $y_\chi$ and to up quarks with strength $y_u$.
For example, $m_\chi \simeq 1.13$~GeV, $m_\phi \simeq 494$~MeV and $\sqrt{y_\chi y_u} \simeq 0.33$ can account for the observed event while remaining allowed by other existing constraints.
\end{abstract}

\maketitle

Despite the remarkable success of the Standard model (SM) of particle physics, many natural phenomena are yet to be resolved \cite{ParticleDataGroup:2026mpi}.
Such shortcomings of SM include dark matter \cite{Cirelli:2024ssz}, neutrino mass \cite{T2K:2011ypd}, the strong CP problem etc. 
For these reasons, the quest for beyond standard model (BSM) physics persists for almost half a century.
The ton-scale dark matter (DM) detectors \cite{PandaX-II:2017hlx,XENON:2020kmp,
LUX-ZEPLIN:2018poe,DEAP:2019yzn}, though originally proposed to identify DM, have served as a powerful strategy to exclude or constrain many such low energy BSM scenarios and strengthened our understanding of the SM neutrino sector.
The recent observation by LUX-ZEPLIN (LZ) experiment has suggested an isolated nuclear recoil event around  
$248\pm 23\pm 23$ keV recoil energy ($T_{\mathcal{N}}$) \cite{LZ:2026axp}.
Interestingly, the origin of such an excess event remains unresolved 
from our known background arising from the SM interactions.
The most dominant background for direct detection experiments is the solar neutrinos, which are accumulated below $T_{\mathcal{N}}\lesssim 10$ keV \cite{LZ:2026axp}.
And above $T_{\mathcal{N}}\gtrsim 100$ keV, events due to
the atmospheric neutrinos, mostly arising from pion decay, are too suppressed.
Since the standard weak interactions can not explain such an event at high recoil energy, this strongly points toward BSM interactions.

The characteristic feature of this unknown event also make its quite distinct from the events arising from BSM scenarios with a continuous falling rate with recoil energy.
To explain the absence of any such BSM excess events in low recoil energy inelastic dark matter (DM) models are proposed \cite{Visinelli:2026kgt,Yamashita:2026ump,DiMauro:2026ldr,Lou:2026idn,Dent:2026bji,deLima:2026shq,Su:2026rwz,Kumar:2026lgi,Chanda:2026wdl}.
In such scenarios, the lighter DM component scatters with a nucleon to produce a heavier dark state.
This poses a lower cutoff on the recoil energy and thus can well explain such a feature. Heavy Higgsino-like DM has also been studied in this context~\cite{Du:2026guj,Yin:2026jnn,Cheung:2026byg,Chatterjee:2026scv}.
Other proposals include an axion portal~\cite{Unwin:2026rdp}, a Majoron portal~\cite{CentellesChulia:2026ats}, a scalar extended $U(1)_{L_\mu-L_\tau}$ model~\cite{Pradhan:2026inq}, and a strongly modulating signal from the high velocity tail of the DM halo~\cite{McCabe:2026crm}.
On the contrary, in this paper we propose 
an alternative explanation of such an excess by BSM interactions of neutrinos.
The existence of neutrino masses is indeed a strong indication of BSM physics, and
many well motivated BSM scenarios propose dark particles to explain the neutrino mass issue.
Neutrinos undergoing elastic scattering (e.g. $\nu \nuc \to \nu \nuc$) through any BSM interactions can in principle induce excess events \cite{AristizabalSierra:2020edu,Brdar:2026ukx}.
However, such scatterings exhibit higher recoil events in the lower $T_\nuc$ and the required interaction rate to produce one event in high $T_\nuc\sim 250$ keV is ruled out from low energy ($T_\nuc\lesssim 30$ keV) observations \cite{DeRomeri:2022twg, Ahmed:2026knn, Chattaraj:2026fxn}.

\begin{figure}[ht!]
    \centering
    \includegraphics[width=1\linewidth]{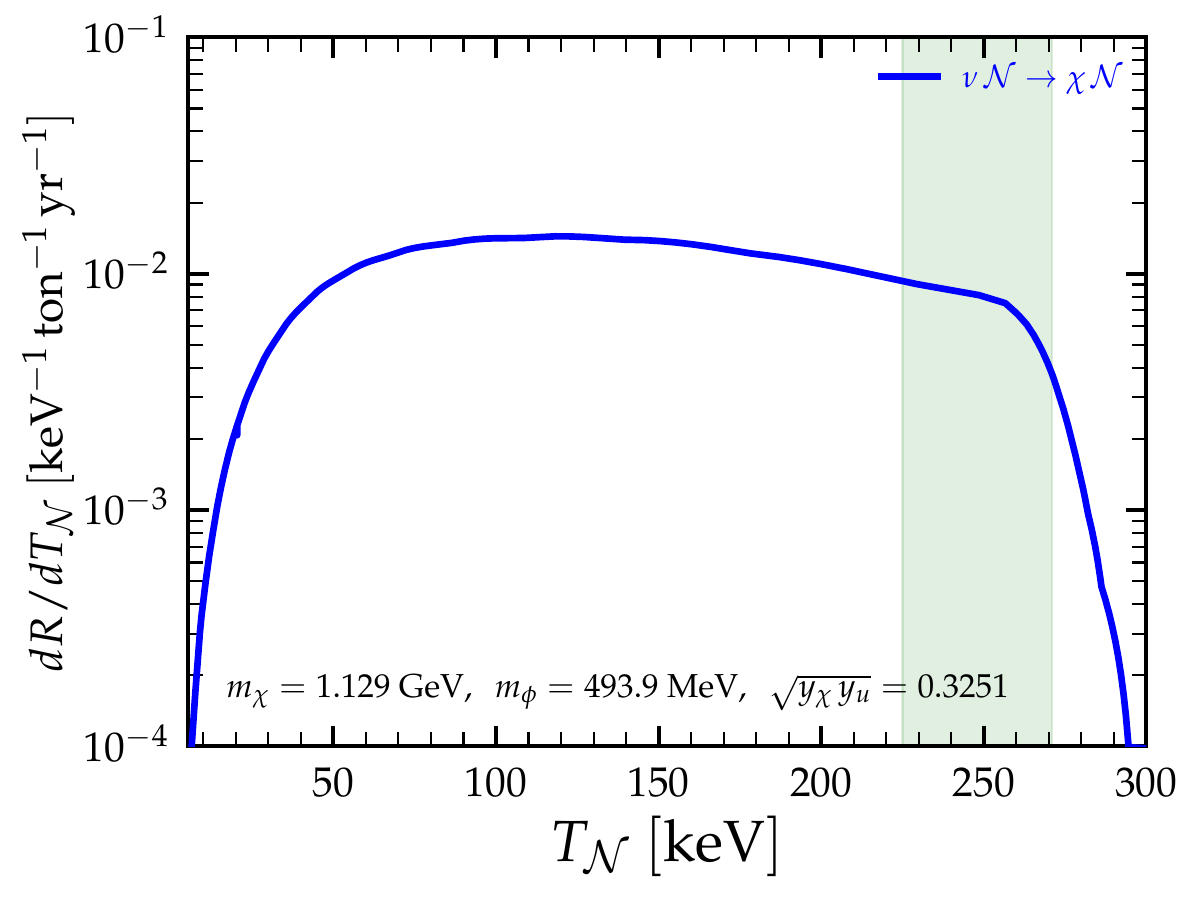}
    \caption{Differential nuclear recoil rate at LZ from atmospheric neutrino up-scattering, $\nu\nuc\to\chi\nuc$, through a pseudoscalar mediator, for the benchmark $m_\chi = 1.129$~GeV, $m_\phi = 493.9$~MeV and $\sqrt{y_\chi y_u} = 0.3251$.
    The green band marks the signal window, $225\text{--}271$~keV, which contains the event observed at $T_\nuc = 248$~keV~\cite{LZ:2026axp}.}
\label{fig:events}
\end{figure}

In this work we consider, the upscattering of $\nu$ to a heavier dark particle $\chi$ with mass $m_\chi$ (e.g. $\nu \nuc \to \chi \nuc$).
Such a scenario is well explored for neutrino scattering events \cite{Brdar:2018qqj,Chen:2021uuw, Candela:2024ljb} and also in light DM absorption searches where $\chi$ can deposit its whole mass-energy to generate the recoil energy of the target \cite{Dror:2019onn,Dror:2019dib,Dror:2020czw}.
However, for our scenario $\chi$ can be any generic BSM particle, not necessarily DM. 
The striking feature of such $\nu$ up-scattering to massive state is that, only neutrinos with energy $\gtrsim m_\chi$ can undergo such scattering and produce the recoil.
Hence it can potentially explain the peaked event at $T_\nuc \sim 250$ keV due to the atmospheric $\nu$ up-scattering without jeopardizing solar $\nu$ scattering events.

The kinematic threshold alone, however, does not guarantee an isolated event. Above the recoil threshold, the shape of the spectrum is set by the Lorentz structure of the interaction.
Since $T_\nuc^{\rm min}$ decreases with increasing $E_\nu$, the atmospheric neutrinos energetic enough to produce $\chi$ at $T_\nuc\simeq 248$~keV also produce recoil events over the entire low recoil window, $5.4$-$225$~keV. So that the ratio of events in this window to those in the signal window, $225$-$271$~keV, is controlled by the $T_\nuc$ dependence of the cross section.
For spin-independent (SI) interactions, i.e. scalar and vector mediators, the nucleon-level vertices add coherently over the nucleus.
At leading order in $1/m_N$, they are also momentum independent.
For a scalar interaction, $\overline{u}_N u_N \simeq 2m_N$, and for a vector, $\overline{u}_N\gamma^\mu u_N \simeq 2m_N\,\delta^{\mu 0}$.
Here $u_N$ is the nucleon spinor, $N=p,n$, and $m_N$ the nucleon mass.
The remaining $T_\nuc$ dependence has three sources.
The kinematic factors are nearly flat for $m_\chi^2 \gg 2m_\nuc T_\nuc$.
The mediator propagator falls with increasing $T_\nuc$.
The nuclear form factor of xenon has its first zero near $T_\nuc\simeq 90$~keV and strongly suppresses the rate in the signal window.
All three favour low recoil energies, so that requiring one event in the signal window overpopulates the low recoil window.
Among spin-dependent (SD) interactions, the axial-vector vertex is likewise momentum independent at leading order, $\overline{u}_N\gamma^i\gamma_5 u_N \simeq 2m_N\,\sigma_N^i$.
It therefore yields the same spectral shape as the SI case, without the coherent enhancement, and suffers from the same problem.
However, the pseudoscalar vertex is different. It connects the upper and lower components of the nucleon spinor, $\overline{u}_N i\gamma_5 u_N \simeq -i\,\boldsymbol{\sigma}_N\cdot\mathbf{q}$, with $\mathbf{q}$ the three momentum transfer, and therefore vanishes in the limit of zero momentum transfer.
The cross section then acquires the additional factor $\qt^2/4m_N^2 = m_\nuc T_\nuc/2m_N^2$, which grows linearly with $T_\nuc$.
Compared with SI and axial-vector interactions, higher recoil energies are therefore enhanced relative to lower ones.
This tilts the spectrum towards high recoil energies, and we therefore consider a pseudoscalar mediator in the following.

The quark couplings of a pseudoscalar mediator cannot be flavour universal.
The nucleon matrix elements of the operators $\overline{q}\, i\gamma_5 q$, with $q=u,d,s$, follow from the anomalous axial Ward identity~\cite{Cheng:2012qr,Bishara:2017pfq},
\begin{equation}
\begin{split}
    \langle N|\, m_q\, \overline{q}\, i\gamma_5 q\, |N\rangle
    &= m_N\, \overline{u}_N\, i\gamma_5\, u_N \\
    &\times
    \Bigg[G_q^N(\qt^2) -
    \bar m \sum_{q'=u,d,s}
    \dfrac{G_{q'}^N(\qt^2)}{m_{q'}}\Bigg],
    \label{eq:ward}
\end{split}
\end{equation}
where $m_q$ are the light quark masses, $\bar m = (1/m_u+1/m_d+1/m_s)^{-1}$, and the second term is the gluonic contribution from the axial anomaly, common to all flavours.
The functions $G_q^N(\qt^2)$ contain the momentum dependence from the $\pi$ and $\eta$ poles,
\begin{equation}
    G_q^N(\qt^2) = \Delta_q^N - \sum_{M=\pi,\eta} a_{q,M}^N\, \dfrac{\qt^2}{\qt^2+m_M^2},
\end{equation}
where $\Delta_q^N$ are the quark spin fractions of the nucleon.
Assuming isospin symmetry, we adopt~\cite{Lin:2018obj, FlavourLatticeAveragingGroup:2019iem, DelNobile:2021wmp}
\begin{equation*}
    \Delta_u^p=\Delta_d^n=0.777,~
    \Delta_d^p=\Delta_u^n=-0.438,~ \Delta_s^p=\Delta_s^n=-0.053.
\end{equation*}
The pole coefficients are $a_{u,\pi}^p=-a_{d,\pi}^p=-a_{u,\pi}^n=a_{d,\pi}^n=g_A^{(3)}/2$, $a_{u,\eta}^N=a_{d,\eta}^N=-a_{s,\eta}^N/2=g_A^{(8)}/6$, $a_{s,\pi}^N=0$. Here $g_A^{(3)}=\Delta_u^p-\Delta_d^p$ and $g_A^{(8)}=\Delta_u^p+\Delta_d^p-2\Delta_s^p$ are the isovector and octet axial charges of the nucleon, respectively.
For a flavour universal coupling, $y_u=y_d=y_s$, the sum over flavours gives $\sum_q \bar m/m_q = 1$, and the two terms in Eq.~\eqref{eq:ward} cancel exactly at any momentum transfer.
The isosinglet pseudoscalar density therefore does not couple to nucleons in this framework, and a non-vanishing nucleon coupling requires flavour non-universal couplings.
As a minimal choice, we take the mediator to couple only to up quarks; a coupling only to down quarks gives a nucleon coupling of the similar size.

To realize such a scenario we consider the following BSM Lagrangian,
\begin{equation}
\label{eq:effective-lagrangian}
     \mathcal{L}_{\rm BSM} \supset y_\chi\, \phi\, \overline{\chi}\, i\gamma_5 P_L \nu_\ell + \mathrm{H.c.} + y_u\, \phi\, \overline{u}\, i\gamma_5 u ,
\end{equation}
where $\phi$ is a pseudoscalar mediator with mass $m_\phi$, and $y_\chi$ ($y_u$) denotes its coupling to $\chi$ (up quarks).

The differential cross section of $\nu\nuc\to\chi\nuc$ scattering reads (see Appendix~\ref{app:cross-section} for the derivation)
\begin{eqnarray}
    \dfrac{\d \sigma_{\nu \nuc}}{\d T_\nuc} (E_\nu, T_\nuc) &=& \dfrac{y_\chi^2\, m_\nuc}{4\pi \left(m_\phi^2 + 2m_\nuc T_\nuc\right)^2}\, S_L(\qt^2) \nonumber\\
    && \times \dfrac{m_\nuc T_\nuc}{2 m_N^2} \left(\dfrac{m_\nuc T_\nuc}{E_\nu^2} + \dfrac{m_\chi^2}{2 E_\nu^2}\right), \label{eq:cross-section-pseudoscalar}
\end{eqnarray}
where $m_\nuc$ is the nuclear mass and $\qt^2 = 2 m_\nuc T_\nuc$.
The nuclear physics enters through the longitudinal spin structure factor~\cite{Fitzpatrick:2012ix,Anand:2013yka}
\begin{equation}
    S_L(\qt^2) = \dfrac{4\pi}{2J_\nuc+1} \sum_{N,N'=p,n} a_N\, a_{N'}\, W_{\Sigma''}^{NN'}(y),
\end{equation}
with $y = (\qt b)^2/4$.
Here $J_\nuc$ is the nuclear spin, $b=\sqrt{41.467/(45A^{-1/3}-25A^{-2/3})}$~fm is the harmonic-oscillator length, and $W_{\Sigma''}^{NN'}$ are the $\Sigma''$ nuclear responses, which we compute with the \texttt{DMFormFactor} package~\cite{Anand:2013yka}.
Only $\Sigma''$ contributes since $\boldsymbol{\sigma}_N\cdot\mathbf{q}$ is purely longitudinal.
As the response vanishes for spin-zero nuclei, only $^{129}$Xe ($J_\nuc=1/2$, abundance $\eta=26.4\%$) and $^{131}$Xe ($J_\nuc=3/2$, $\eta=21.29\%$) contribute.

Using Eq.~\eqref{eq:ward}, the nucleon couplings $a_N$ that encode the mapping of quark level interactions to the nucleon read~\cite{Cheng:2012qr,Bishara:2017pfq},
\begin{equation}
    a_N = y_u\, \dfrac{m_N}{m_u} \left[G_u^N(\qt^2) - \bar m \sum_{q=u,d,s} \dfrac{G_q^N(\qt^2)}{m_q}\right].
\end{equation}
The pion pole reduces the isovector part of $G_q^N$ by $\sim 77\%$ at $T_\nuc=250$~keV, so the momentum dependence is not a small correction.

\begin{figure}[ht!]
    \centering
    \includegraphics[width=1\linewidth]{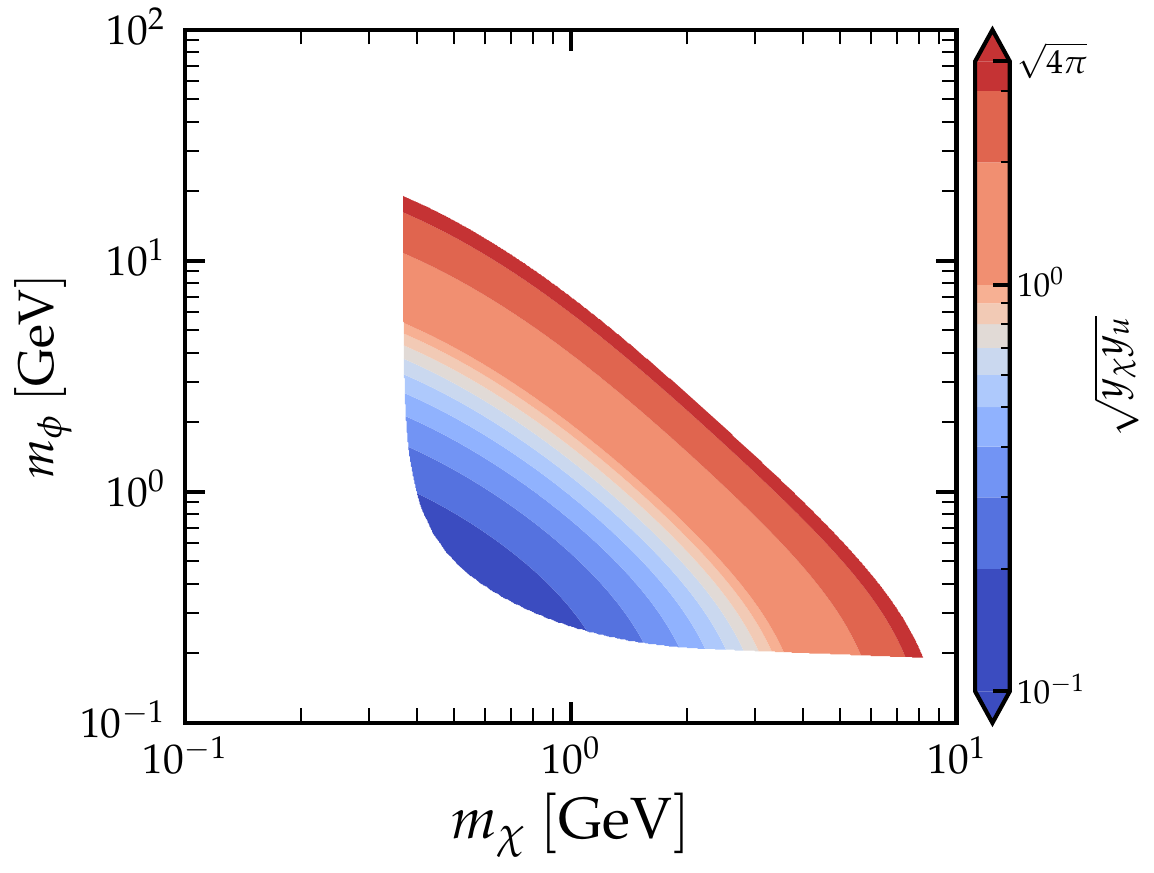}
    \caption{Region of the $(m_\chi,\, m_\phi)$ plane in which atmospheric neutrino up-scattering, $\nu\nuc\to\chi\nuc$, through a pseudoscalar mediator accounts for the LZ event.
    At each point, the coupling $\sqrt{y_\chi y_u}$ (colour scale) is fixed by requiring $R_{\rm I}=1$ event in $225\text{--}271$~keV.
    Only points with $R_{\rm II}<10$ events in $5.4\text{-}225$~keV and $\sqrt{y_\chi y_u}<\sqrt{4\pi}$ are shown.
    The region closes at small $m_\chi$ and small $m_\phi$, where the spectrum overpopulates the low recoil window, giving $R_{\rm II}>10$.
    Towards large $m_\chi$ and $m_\phi$, the suppressed atmospheric flux above $E_\nu^{\rm min}$ and the heavy mediator suppression require $\sqrt{y_\chi y_u}>\sqrt{4\pi}$.}
\label{fig:param-space}
\end{figure}

Compared to the elastic scattering of neutrino, the expected spectra of such up-scattering has a strong dependence on the produced BSM particle $\chi$.
The minimum (maximum) recoil energy is given by,
\begin{eqnarray}
      T_{\mathcal{N}}^{\rm min/max}  &=& \dfrac{1}{2m_{\nuc}\left(2E_{\nu}^{\mathrm{max}} + m_{\nuc}\right)} \times \nonumber \\
      &&\Bigg[2m_{\nuc}{(E_{\nu}^{\mathrm{max}})}^{2} - m_\chi^2\left(E_{\nu}^{\mathrm{max}} + m_{\nuc}\right) \nonumber \\
      &&\pm E_{\nu}^{\mathrm{max}}\bigg(4m_{\nuc}^2 {(E_{\nu}^\mathrm{max})}^{2} 
      \nonumber \\
      && - 4m_{\nuc} m_\chi^2 \left(E_{\nu}^{\mathrm{max}} + m_{\nuc}\right) + m_\chi^4\bigg)^{1/2}\Bigg].
\end{eqnarray}
$E_{\nu}^{\mathrm{max}}$ denotes the maximum energy of each component.
For a given flux component, up-scattering is kinematically allowed only if $E_\nu^{\rm max} \gtrsim m_\chi$.
Solar and DSNB neutrinos, with $E_\nu^{\rm max} \lesssim 100$~MeV, therefore cannot produce a GeV-scale $\chi$ at all, and contribute only through elastic scatterings.
The atmospheric flux, in contrast, extends to TeV energies, far above $m_\chi$.
Atmospheric neutrinos can thus populate the entire recoil window through up-scattering, and the shape of the spectrum is set by the Lorentz structure of the interaction.
The recoil spectrum is  given by,
    \begin{eqnarray}
   \frac{\d R}{\d T_\nuc} &=& t_{\rm exp}\mathcal{E}(T_\mathcal{N})N_T~   \int_{E_{\nu}^{\mathrm{min}}}^{E_{\nu}^{\mathrm{max}}} \mathrm{d}E_\nu ~\sum_{\ell} ~\dfrac{\mathrm{d}\Phi_{\nu_{\ell}}}{\mathrm{d} E_\nu}~ \dfrac{\mathrm{d}\sigma_{\nu\mathcal{N}}}{\mathrm{d}T_\nuc} ~ \nonumber \\
      && \times
   \Theta(T_\nuc-T_\nuc^{\mathrm{min}})~\Theta(T_\nuc^{\mathrm{max}}-T_\nuc),
\end{eqnarray}
where we consider all astrophysical neutrinos, such as solar, DSNB and atmospheric neutrinos, summed over all flavours.
For the atmospheric component we adopt the spectrum of Ref.~\cite{Vitagliano:2019yzm}, which extends from $\sim 10$~MeV to $\sim 100$~TeV. 
$t_{\rm exp}$ and $N_T$ signify time of exposure and number of target nucleus, while $\mathcal{E}(T_\nuc)$ denotes the total nuclear recoil signal efficiency of LZ after all selection cuts~\cite{LZ:2026axp}.
$E_{\nu}^{\mathrm{min}}$ describes the minimum energy required to up-scatter, 
\begin{eqnarray}
    E_{\nu}^{\mathrm{min}} (T_\nuc) &=& \dfrac{1}{2} \left(T_\nuc + \sqrt{2 m_\nuc T_\nuc + (T_\nuc)^2}\right)\nonumber \\
      && \times
    \left(1 + \dfrac{m_\chi^2}{2 m_\nuc T_\nuc}\right).
\end{eqnarray}
The event rate for our region of interest is given by
\begin{eqnarray}
    R_{I}=~\int_{225{\rm ~keV}}^{271{\rm ~keV}} \mathrm{d}T_\nuc~ \frac{\d R}{\d T_\nuc}.
\end{eqnarray}
To explain the observed event, we require
$R_{I}=1$.
Though the kinematic requirement of the up-scattering process prevents excess events from solar and DSNB neutrinos, the high energy neutrinos ($E_\nu\gtrsim m_\chi$) in the tail of the atmospheric $\nu$ flux can generate non-zero events at low $T_\nuc$.
To this end, the momentum dependent operator considered here is crucial.
The $q^2$ suppression arising from the pseudo-scalar mediator suppresses the events at low $T_\nuc$ as mentioned earlier.

Hence, besides the isolated event at $\sim 250$ keV, one also has to make sure that the tail of the BSM event spectrum in lower recoil energies 
remain consistent with the observed events.
The collaboration has reported a fitted background $R_{\rm BKG}=1713\pm 39$ and observed counts $R_{\rm obs}=1710$ \cite{LZ:2026axp}.
To prevent any excess events in other energy ranges, we also impose the criterion,
\begin{eqnarray}
    R_{II}=~\int_{5.4{\rm ~keV}}^{225{\rm ~keV}} \mathrm{d}T_\nuc~ \frac{\d R}{\d T_\nuc}< 10.
\end{eqnarray}
which is a conservative choice, smaller than the uncertainties in the background \cite{LZ:2026axp}.
This crude cut on the excess events in lower recoil energy ensures that we BSM events in that range remain negligible. 

The predicted differential recoil rate for a representative benchmark, $m_\chi = 1.129$~GeV, $m_\phi = 493.9$~MeV and $\sqrt{y_\chi y_u} = 0.3251$, is shown in Fig.~\ref{fig:events}.
The rate rises steeply above the analysis threshold and remains nearly flat up to the signal window.
This flatness results from the $\qt^2$ factor of the pseudoscalar vertex, which compensates the fall off of the nuclear spin response with increasing $T_\nuc$.
Above $T_\nuc \simeq 250$~keV the rate drops, following the LZ efficiency.
For this benchmark, the up-scattering yields $R_{\rm I}=1$ event in the signal window, while the events spread over the low recoil window remain below our $R_{\rm II}$ criterion.

The region of parameter space that accounts for the observed event is presented in Fig.~\ref{fig:param-space}.
It spans $0.37~{\rm GeV} \lesssim m_\chi \lesssim 8.2$~GeV and $0.19~{\rm GeV} \lesssim m_\phi \lesssim 19$~GeV, with couplings $0.13 \lesssim \sqrt{y_\chi y_u} < \sqrt{4\pi}$.
The region is closed from all sides.
For small $m_\chi$ or small $m_\phi$, the spectrum overpopulates the low recoil window, violating our $R_{\rm II}$ criterion.
Towards large $m_\chi$, $E_\nu^{\rm min}$ increases and the atmospheric flux above that energy decreases, while towards large $m_\phi$ the propagator suppresses the cross section.
In both cases, one event in the signal window would require $\sqrt{y_\chi y_u} > \sqrt{4\pi}$.
Within the allowed region, the required coupling grows with both $m_\chi$ and $m_\phi$, and remains consistent with the existing constraints discussed below.

High energy colliders like LHC  so far set the strongest constraint on such BSM operator ($y_\chi y_q/m_\phi^2$) through mono-jet signals (i.e. $pp\to \chi+\Bar{\nu}$+jet) \cite{ATLAS:2020uiq}.
However, the center of mass energy (COM) considered in the LHC analysis was performed $\sqrt{s}=$13\,TeV with a typical momentum transfer at the parton level $Q_{\rm tr}\gtrsim 800$ GeV \cite{Busoni:2014sya} and thus models featuring a mediator $m_\phi\lesssim 800$\,GeV can not be realized in the EFT approach. On the other hand, light mediators with mass $\ll Q_{\rm tr}$ will be produced on-shell and
Meson decay constraints from Belle II, NA62 may exclude some part of the parameter space from meson decay observations \cite{Liu:2025lbw,Mescia:2026xju} leaving the parameter space $m_\chi\gtrsim 800$ MeV unconstrained.
BaBar constraint from $\Upsilon$ 
decay does not apply since $\phi$ couples to only $u$ quark.
On the other hand, radiative $Z$ decays \cite{Davidson:2003ha} can not probe the parameter space for $m_{\phi}\ll M_{Z} $  where on-shell $\phi$ gets produced, leaving the $ \phi-\chi$ interaction unconstrained. 
Hence, the existing limits from these scenarios on the EFT operator can not be translated to our framework, requiring a dedicated analysis as a complementary probe of our proposed BSM setup.
Astrophysical constraints like supernova cooling \cite{Lin:2025mez} %
becomes weak for $m_\chi\gtrsim 300$ MeV.

Recently, Ref.~\cite{Brdar:2026ukx} pointed out that a single step up-scattering of atmospheric neutrinos, $\nu\nuc\to\chi\nuc$, also predicts neutral current like events in large neutrino detectors. On lighter nuclei, such as oxygen and carbon, the same process gives larger recoil energies than on xenon, and no such excess has been observed. In our scenario this argument is weakened by the spin structure of the interaction. The pseudoscalar coupling enters only through $S_L$, which vanishes for spin-zero nuclei. The main targets of water Cherenkov and liquid scintillator detectors, $^{16}$O and $^{12}$C, have $J_\nuc=0$ and do not contribute. Only the rare isotopes $^{13}$C and $^{17}$O, with natural abundances of about $1.1\%$ and $0.04\%$, can scatter through this interaction. For our pseudoscalar interaction, the constraint from these detectors is therefore relaxed compared with spin independent up-scattering.

In this work, we demonstrate that atmospheric neutrino conversion to a heavy dark fermion through a BSM scalar mediator can in principle explain the single isolated nuclear recoil event observed by LZ.
The heavy final state $\chi$ with mass $\sim$GeV, imposes a kinematic cutoff preventing excess events from solar and DSNB $\nu$ at low $T_\nuc$ window.
On the other hand, the $q^2$ suppression arising from the pseudo-scalar mediator suppresses the events at low $T_\nuc$ window generated by the high-energy tail of atmospheric $\nu$ flux, 
mimicking an isolated BSM event.
Indeed, a single event is insufficient to pin down uniquely the microscopic BSM physics. 
In the near future, other direct search facilities with extremely low backgrounds can offer an ideal test of this explanation.
Larger detectors like KamLAND \cite{KamLAND:2021gvi} and Borexino \cite{BOREXINO:2021efb} can also probe the best-fit parameter space identified here, which we keep for our future work.

{\it Acknowledgement.} 
We thank Dilip K. Ghosh, Sanjay Mondal, Jo\~ao Paulo Pinheiro and Oleg Titov, Ayan Chattaraj, Dimitrios K. Papoulias for helpful discussions and useful comments on the manuscript.
SJ thanks Xiaoyong Chu and ICTP-AP, Beijing, for arranging an academic visit where this work was accomplished.
SJ acknowledges financial support from the National Natural Science Foundation of China (12425506 and 12375101) and State Key Laboratory of Dark Matter Physics.

\appendix

\section{Derivation of the $\nu\nuc\to\chi\nuc$ cross section}
\label{app:cross-section}

In this appendix we derive the differential cross section of Eq.~\eqref{eq:cross-section-pseudoscalar}, starting from the Lagrangian of Eq.~\eqref{eq:effective-lagrangian}.
We denote the momenta by $\nu(p_1)+\nuc(p_2)\to\chi(k_1)+\nuc(k_2)$ and define the four-momentum transfer $q=k_2-p_2=p_1-k_1$.
In the laboratory frame the nucleus is initially at rest, $p_2=(m_\nuc,\mathbf{0})$, and
\begin{equation}
    t \equiv q^2 = (k_2-p_2)^2 = -2m_\nuc T_\nuc, \qquad \qt^2 = 2m_\nuc T_\nuc ,
    \label{eq:app-t}
\end{equation}
where $\qt^2=-t$ is the squared three-momentum transfer in the Breit frame of the nucleus.
We use the relativistic normalization $\langle \mathbf{p}'|\mathbf{p}\rangle = 2E\,(2\pi)^3\delta^3(\mathbf{p}-\mathbf{p}')$ for all external states.

\subsection{From quarks to nucleons}
\label{app:quark-nucleon}

The mediator couples to the up quark pseudoscalar density.
Its nucleon matrix element follows from Eq.~\eqref{eq:ward},
\begin{eqnarray}
    && \langle N(k')|\, y_u\,\overline{u}\, i\gamma_5 u\, |N(k)\rangle \nonumber\\
    &&\quad = \dfrac{y_u}{m_u}\,\langle N(k')|\, m_u\,\overline{u}\, i\gamma_5 u\, |N(k)\rangle \nonumber\\
    &&\quad = a_N(\qt^2)\; \overline{u}_N(k')\, i\gamma_5\, u_N(k),
    \label{eq:app-nucleon-me}
\end{eqnarray}
with $N=p,n$ and
\begin{equation}
    a_N(\qt^2) = y_u\, \dfrac{m_N}{m_u} \left[G_u^N(\qt^2) - \bar m \sum_{q=u,d,s} \dfrac{G_q^N(\qt^2)}{m_q}\right].
    \label{eq:app-aN}
\end{equation}
The quark-level interaction is therefore equivalent to the nucleon-level Lagrangian
\begin{equation}
    \mathcal{L}_N = \phi \sum_{N=p,n} a_N\, \overline{N}\, i\gamma_5 N ,
\end{equation}
where $\overline{N}$ and $N$ denote the nucleon fields.

\subsection{Scattering amplitude}
\label{app:amplitude}

At tree level, the process proceeds through $t$-channel exchange of $\phi$.
The amplitude reads
\begin{eqnarray}
    \mathcal{M} &=& \dfrac{y_\chi}{t-m_\phi^2}\; \big[\overline{u}_\chi(k_1)\, i\gamma_5 P_L\, u_\nu(p_1)\big] \nonumber\\
    && \times\, \langle \nuc(k_2)|\, J_P(0)\, |\nuc(p_2)\rangle ,
    \label{eq:app-amplitude}
\end{eqnarray}
up to an overall phase, where
\begin{equation}
    J_P(x) = \sum_{N=p,n} a_N\, \overline{N}(x)\, i\gamma_5 N(x)
\end{equation}
is the nucleon pseudoscalar current.
All nuclear physics is contained in the matrix element $\langle \nuc'|J_P|\nuc\rangle$, which we evaluate in Secs.~\ref{app:nucleon-reduction} and~\ref{app:nuclear}.

\subsection{Leptonic part}
\label{app:lepton}

The leptonic part of the amplitude square is given by,
\begin{eqnarray}
    L &\equiv& \sum_{\rm spins} \big|\overline{u}_\chi(k_1)\, i\gamma_5 P_L\, u_\nu(p_1)\big|^2 \nonumber\\
    &=& 2\, p_1\cdot k_1 .
\end{eqnarray}
With $m_\nu=0$, Eq.~\eqref{eq:app-t} gives $t=m_\chi^2-2\,p_1\cdot k_1$, so that
\begin{equation}
    L = m_\chi^2 + 2m_\nuc T_\nuc .
    \label{eq:app-lepton}
\end{equation}
The term $m_\chi^2$ reflects the chirality flip at the leptonic vertex, which requires a massive final state.

\subsection{Non-relativistic reduction of the nucleon vertex}
\label{app:nucleon-reduction}

In the Dirac representation,
\begin{eqnarray}
    \gamma^0\gamma_5 &=& \begin{pmatrix} 0 & 1 \\ -1 & 0 \end{pmatrix}, \nonumber\\
    u_N(k) &=& \sqrt{E+m_N}\begin{pmatrix} \xi \\[2pt] \dfrac{\boldsymbol{\sigma}\cdot\mathbf{k}}{E+m_N}\,\xi \end{pmatrix},
\end{eqnarray}
where $\xi$ is a two-component Pauli spinor with $\xi^\dagger\xi=1$.
Since $\gamma^0\gamma_5$ is off-diagonal, the pseudoscalar bilinear connects the upper and lower components,
\begin{equation}
\begin{split}
    \overline{u}_N(k')\, i\gamma_5\, u_N(k)
    &= i\sqrt{(E'+m_N)(E+m_N)} \\
    &\times \xi'^\dagger\bigg[\dfrac{\boldsymbol{\sigma}\cdot\mathbf{k}}{E+m_N}
      - \dfrac{\boldsymbol{\sigma}\cdot\mathbf{k}'}{E'+m_N}\bigg]\xi .
\end{split}
\end{equation}
For nucleons bound in the nucleus, $E\simeq E'\simeq m_N$ up to corrections of order $|\mathbf{k}|^2/m_N^2$, which gives
\begin{equation}
    \overline{u}_N(k')\, i\gamma_5\, u_N(k) \simeq -i\, \xi'^\dagger\, \boldsymbol{\sigma}\cdot\mathbf{q}\; \xi ,
    \qquad \mathbf{q}=\mathbf{k}'-\mathbf{k} .
    \label{eq:app-NR}
\end{equation}
The prefactor $\sqrt{(E'+m_N)(E+m_N)}\simeq 2m_N$ cancels against $1/(E+m_N)$ in the lower components.
Unlike the scalar bilinear, $\overline{u}_N u_N\simeq 2m_N$, the pseudoscalar bilinear therefore carries no factor of $2m_N$ and vanishes at zero momentum transfer.

\subsection{Nuclear matrix element}
\label{app:nuclear}

The nucleus is described by a non-relativistic wave function $\Psi$, normalized to unity.
Relativistic and non-relativistic states are related by $|\nuc\rangle_{\rm rel}=\sqrt{2E_\nuc}\,|\nuc\rangle_{\rm NR}$.
In the impulse approximation, $J_P$ acts on one nucleon at a time, and each nucleon contributes its bilinear divided by its own normalization $2m_N$.
In the non-relativistic limit, $E_\nuc\simeq m_\nuc$, this gives
\begin{eqnarray}
    && \langle \nuc'|\, J_P\, |\nuc\rangle \nonumber\\
    &&\quad \simeq 2m_\nuc \Big\langle \Psi' \Big| \sum_{i=1}^{A} a_{N_i}\, \dfrac{\big[\overline{u}_N\, i\gamma_5\, u_N\big]_i}{2m_N}\, e^{i\mathbf{q}\cdot\mathbf{r}_i} \Big| \Psi \Big\rangle \nonumber\\
    &&\quad = -i\, \dfrac{m_\nuc}{m_N}\, \qt \nonumber\\
    &&\qquad \times \Big\langle \Psi' \Big| \sum_{i=1}^{A} a_{N_i}\, \boldsymbol{\sigma}_i\cdot\hat{\mathbf{q}}\; e^{i\mathbf{q}\cdot\mathbf{r}_i} \Big| \Psi \Big\rangle ,
    \label{eq:app-nuclear-me}
\end{eqnarray}
where we used Eq.~\eqref{eq:app-NR}, $\mathbf{r}_i$ is the position of nucleon $i$, and $\hat{\mathbf{q}}=\mathbf{q}/\qt$.
The factor $m_\nuc/m_N$ is the ratio of the nuclear and nucleon state normalizations.

The operator $\boldsymbol{\sigma}\cdot\hat{\mathbf{q}}\,e^{i\mathbf{q}\cdot\mathbf{r}}$ is purely longitudinal.
Choosing $\hat{\mathbf{q}}=\hat{\mathbf{z}}$, its multipole expansion reads~\cite{Fitzpatrick:2012ix,Anand:2013yka}
\begin{equation}
    \boldsymbol{\sigma}\cdot\hat{\mathbf{q}}\; e^{i\mathbf{q}\cdot\mathbf{r}}
    = -i \sum_{J} i^J \sqrt{4\pi(2J+1)}\; \Sigma''_{J0}(\qt\mathbf{r}),
\end{equation}
with
\begin{equation}
    \Sigma''_{JM}(\qt\mathbf{r}) = \boldsymbol{\sigma}\cdot\dfrac{\nabla}{\qt}\, \Big[ j_J(\qt r)\, Y_{JM}(\hat{\mathbf{r}}) \Big] .
\end{equation}
For elastic scattering on a nuclear ground state of definite parity, only odd $J$ contribute.

We average over the initial and sum over the final nuclear spin projections $M,M'$.
With the Wigner--Eckart theorem and the orthogonality of the $3j$ symbols, the cross terms between different $J$ vanish, and
\begin{eqnarray}
    && \dfrac{1}{2J_\nuc+1} \sum_{M,M'} \Big| \Big\langle J_\nuc M' \Big| \sum_{i} a_{N_i} \times\, \boldsymbol{\sigma}_i\cdot\hat{\mathbf{q}}\; e^{i\mathbf{q}\cdot\mathbf{r}_i} \Big| J_\nuc M \Big\rangle \Big|^2 \nonumber\\
    &&\quad = \dfrac{4\pi}{2J_\nuc+1} \sum_{N,N'} a_N\, a_{N'}\, W_{\Sigma''}^{NN'}(y) \nonumber\\
    &&\quad \equiv S_L(\qt^2) ,
    \label{eq:app-SL}
\end{eqnarray}
where $y=(\qt b)^2/4$ and
\begin{equation}
    W_{\Sigma''}^{NN'}(y) = \sum_{J\,{\rm odd}} \langle J_\nuc \| \Sigma''_{J;N} \| J_\nuc \rangle\, \langle J_\nuc \| \Sigma''_{J;N'} \| J_\nuc \rangle
\end{equation}
are the longitudinal spin responses, with $\Sigma''_{J;N}$ the multipole operator summed over the nucleons of type $N$.
Combining Eqs.~\eqref{eq:app-nuclear-me} and~\eqref{eq:app-SL}, the spin-averaged nuclear part of the squared amplitude is
\begin{equation}
    \overline{\big|\langle \nuc'|\, J_P\, |\nuc\rangle\big|^2} = \left(\dfrac{m_\nuc}{m_N}\right)^2 \qt^2\, S_L(\qt^2) .
    \label{eq:app-hadron}
\end{equation}
For spin-zero nuclei $S_L$ vanishes identically, so in xenon only $^{129}$Xe and $^{131}$Xe contribute.

\subsection{Squared amplitude and cross section}
\label{app:cross-section-final}

Inserting Eqs.~\eqref{eq:app-lepton} and~\eqref{eq:app-hadron} into the square of Eq.~\eqref{eq:app-amplitude}, and using $t-m_\phi^2=-(m_\phi^2+2m_\nuc T_\nuc)$, we obtain
\begin{eqnarray}
    \overline{|\mathcal{M}|^2} &=& \dfrac{y_\chi^2\, \big(m_\chi^2 + 2m_\nuc T_\nuc\big)}{\big(m_\phi^2+2m_\nuc T_\nuc\big)^2} \nonumber\\
    && \times\, \dfrac{m_\nuc^2}{m_N^2}\, 2m_\nuc T_\nuc\, S_L(\qt^2) .
    \label{eq:app-M2}
\end{eqnarray}

For the $2\to2$ kinematics, the differential cross section is
\begin{equation}
    \dfrac{\d\sigma}{\d t} = \dfrac{\overline{|\mathcal{M}|^2}}{16\pi\, \lambda(s,0,m_\nuc^2)} ,
\end{equation}
where $\lambda$ is the K\"all\'en function.
With $s=m_\nuc^2+2m_\nuc E_\nu$ in the laboratory frame, $\lambda(s,0,m_\nuc^2)=(s-m_\nuc^2)^2=4m_\nuc^2E_\nu^2$.
Using $\d t = 2m_\nuc\, \d T_\nuc$ from Eq.~\eqref{eq:app-t}, we find
\begin{equation}
    \dfrac{\d\sigma_{\nu\nuc}}{\d T_\nuc} = \dfrac{\overline{|\mathcal{M}|^2}}{32\pi\, m_\nuc E_\nu^2} .
    \label{eq:app-dsdT}
\end{equation}
Substituting Eq.~\eqref{eq:app-M2} gives
\begin{eqnarray}
    \dfrac{\d \sigma_{\nu \nuc}}{\d T_\nuc} (E_\nu, T_\nuc) &=& \dfrac{y_\chi^2\, m_\nuc}{4\pi \left(m_\phi^2 + 2m_\nuc T_\nuc\right)^2}\, S_L(\qt^2) \nonumber\\
    &\times& \dfrac{m_\nuc T_\nuc}{2 m_N^2} \left(\dfrac{m_\nuc T_\nuc}{E_\nu^2} + \dfrac{m_\chi^2}{2 E_\nu^2}\right).
\end{eqnarray}
The factor $m_\nuc T_\nuc/2m_N^2=\qt^2/4m_N^2$ originates from the nucleon vertex of Eq.~\eqref{eq:app-NR}.
It grows linearly with the recoil energy and is responsible for the hard recoil spectrum discussed in the main text.

\subsection{Nuclear responses from \texttt{DMFormFactor}}
\label{app:dmformfactor}

We compute $W_{\Sigma''}^{NN'}$ with the \texttt{DMFormFactor} package~\cite{Anand:2013yka}, 
using its default shell model one-body density matrices and harmonic oscillator length 
$b=\sqrt{41.467/(45A^{-1/3}-25A^{-2/3})}$~fm.
The package returns the nuclear responses in the isospin basis, $W^{\tau\tau'}_{\Sigma''}$ with 
$\tau,\tau'=0,1$, which multiply the isoscalar and isovector couplings
\begin{equation}
    c_0 = c_p + c_n\,, \qquad c_1 = c_p - c_n\,.
    \label{eq:iso_couplings}
\end{equation}
Since the scattering rate depends on the couplings only through the coupling-weighted 
response, the two bases must satisfy
\begin{equation}
    \sum_{N,N'=p,n} c_N\, c_{N'}\, W^{NN'}_{\Sigma''} = \sum_{\tau,\tau'=0,1} c_\tau\, c_{\tau'}\, W^{\tau\tau'}_{\Sigma''}\,.
    \label{eq:basis_invariance}
\end{equation}
Writing Eq.~\eqref{eq:iso_couplings} as $c_\tau=\sum_N A_{\tau N}\,c_N$, with
\begin{equation}
    A = \begin{pmatrix} A_{0p} & A_{0n} \\ A_{1p} & A_{1n} \end{pmatrix}
      = \begin{pmatrix} 1 & 1 \\ 1 & -1 \end{pmatrix},
\end{equation}
substituting into the right hand side of Eq.~\eqref{eq:basis_invariance}, and matching the 
coefficients of $c_N c_{N'}$, the responses transform as
\begin{equation}
    W^{NN'}_{\Sigma''} = \sum_{\tau,\tau'} A_{\tau N}\, A_{\tau' N'}\, W^{\tau\tau'}_{\Sigma''}\,,
    \label{eq:W_transform}
\end{equation}
or explicitly
\begin{eqnarray}
    W^{pp}_{\Sigma''} &=& W^{00}_{\Sigma''}+W^{01}_{\Sigma''}+W^{10}_{\Sigma''}+W^{11}_{\Sigma''}\,, \nonumber\\
    W^{nn}_{\Sigma''} &=& W^{00}_{\Sigma''}-W^{01}_{\Sigma''}-W^{10}_{\Sigma''}+W^{11}_{\Sigma''}\,, \nonumber\\
    W^{pn}_{\Sigma''} &=& W^{00}_{\Sigma''}-W^{01}_{\Sigma''}+W^{10}_{\Sigma''}-W^{11}_{\Sigma''}\,, \nonumber\\
    W^{np}_{\Sigma''} &=& W^{00}_{\Sigma''}+W^{01}_{\Sigma''}-W^{10}_{\Sigma''}-W^{11}_{\Sigma''}\,.
    \label{eq:W_pn_basis}
\end{eqnarray}
For a single (non interfering) 
operator such as $\Sigma''$, the proton-neutron response matrix is symmetric, 
$W^{pn}_{\Sigma''}=W^{np}_{\Sigma''}$, which implies $W^{01}_{\Sigma''}=W^{10}_{\Sigma''}$ and hence
\begin{equation}
    W_{\Sigma''}^{pn} = W_{\Sigma''}^{np} = W_{\Sigma''}^{00} - W_{\Sigma''}^{11}\,.
\end{equation}

\bibliography{ref}
\bibliographystyle{JHEP}

\end{document}